\documentclass[a4paper,11pt]{article}
\usepackage[utf8]{inputenc}
\usepackage[greek,english]{babel}
\usepackage{graphicx}
\usepackage{textcomp}
\usepackage{color}
\usepackage{amssymb}
\usepackage{amsmath}
\usepackage{hyperref}
\usepackage{times}
\usepackage[authoryear,round]{natbib}
\usepackage{multirow}
\usepackage{arydshln}

\title{On The Role Of Simplicity In Science}
\author{Luigi Scorzato \\ ECT* - Trento, Italy}
\date{24 March 2012}

\begin{document}
\maketitle

\begin{center} {\large \bf Abstract} \end{center}
\begin{quote}

Simple assumptions represent a decisive reason to prefer one theory to another in everyday scientific praxis.  But
this praxis has little philosophical justification, since there exist many notions of simplicity, and those that
can be defined precisely strongly depend on the language in which the theory is formulated.  The language
dependence is a natural feature---to some extent---but it is also believed to be a fatal problem, because,
according to a common general argument, the simplicity of a theory is always trivial in a suitably chosen language.
But, this {\em trivialization argument} is typically either applied to toy-models of scientific theories or applied
with little regard for the empirical content of the theory.  This paper shows that {\em the trivialization argument
  fails, when one considers realistic theories and requires their empirical content to be preserved}.  In fact, the
concepts that enable a very simple formulation, are not necessarily measurable, in general.  Moreover, the
inspection of a theory describing a {\em chaotic billiard} shows that precisely those concepts that naturally make
the theory extremely simple are provably not measurable.  This suggests that---whenever a theory possesses
sufficiently complex consequences---the constraint of measurability prevents too simple formulations in any
language.  This explains why the scientists often regard their assessments of simplicity as largely unambiguous.
In order to reveal a cultural bias in the scientists' assessment, one should explicitly identify different
characterizations of simplicity of the assumptions that lead to different theory selections.  General arguments are
not sufficient.

\end{quote}

\section{Introduction}

In order to appreciate the important role of the idea of simplicity, it is worth reviewing one of the most
challenging open questions, concerning our understanding of science.

Most scientists believe that the main goal of their work, namely that of finding {\em better theories} than those
representing the state of the art, is well defined and the criteria for success are reliable and do not depend on
the particular culture dominating the scientific community to which they belong.  Although the scientists are not
immune to disputes, even bitter, the latter occur on rather minor issues, compared to the common grounds that unite
the scientific community.  In particular, it is certainly not true that {\em for any} two competing theories, all
scientists agree on which one is better, but {\em there do exist many and significant} pairs of theories where all
scientists agree that one is unambiguously better than the other.  Moreover, many issues, that divided the
scientists in the past, are now fully settled.

This high level of convergence begs for an explanation.  A challenge for philosophy of science is to understand
whether the standards that scientists perceive as reliable are actually well-grounded on unambiguous cognitive
values---and, if so, identify such values---or, alternatively, identify the cultural bias in the scientists'
assessments, and show how different---but in principle equally admissible---cultural prejudices would lead to
different assessments, {\em even in questions where the scientists unanimously agree}.  In order to justify the
first conclusion, one should identify {\em general} and {\em durable} criteria for comparing two scientific
theories, which are based on {\em unambiguous cognitive values}.  Moreover, the criteria should be {\em usable in
  practice} to select among real scientific theories.

Incommensurability \citep{Kuhn1, Bird-SEP} is sometimes believed to be a stumbling block undermining any {\em
  general} criterion for the comparison of scientific theories.  The alternative is to acknowledge the necessity of
irreducibly different criteria of theory appraisal, for different scientific domains.  This is a favored view among
many philosophers, which is also not strongly opposed by scientists, who have limited authority to judge beyond
their own disciplines (and might even be seduced by the shortsighted illusion that granting the full responsibility
of the judgment to {\em experts} is good for them).  But, it should be clear that the lack of a {\em general}
criterion is ultimately equivalent to no reliable criterion at all, with the consequence that {\em anything goes}
\citep{Feyerabend}.  In fact, it is not uncommon that a dispute over the scientific value of a method, or of a
theory, results in the foundation of a new discipline, with its own alleged scientific standards and experts.  If
we deny any general standard in science, we have to accept such practices as perfectly justified ways of doing
science.

A general criterion for the comparison of different scientific theories---which has also an obvious cognitive
value---is empirical adequacy\footnote{In this paper, the precise definition of {\em empirical adequacy} does not
  play any important role.  Only the concept of {\em empirical equivalence} matters, and it is defined later.}, but
it cannot be the only one.  In fact, empirical adequacy can be easily improved by introducing ad-hoc assumptions
and building more and more complex theories that adapt themselves to the data, without producing any cognitive
advantage.  It has been argued \citep{Sokal1C} that there is often just one theory---at best---that is compatible
with the data and it is not {\em crazy} (such as theories that might be motivated by solipsism, radical skepticism
and other implausible speculations).  This suggests that empirical adequacy should be sufficient for theory
appraisal, provided that one excludes crazy theories.  But unfortunately, there is no sharp distinction between
crazy and non-crazy theories.  How many ad-hoc assumptions are we willing to accept before declaring a theory
crazy?  For example, a full class of gravitational theories within the parametrized post Newtonian (ppN) formalism
\citep{lrr-2006-3} are in agreement with the experimental data as precisely as general relativity (GR).  But GR is
still unanimously regarded as unambiguously better than most of those theories\footnote{This does {\em not} refer
  to those ppN theories that are in {\em better} agreement with some experimental data than GR, like those used to
  model Dark Matter.  These do represent interesting alternatives, and are the reason why the ppN formalism is
  studied.}.  These are not crazy theories at all, but we should nevertheless be able to tell precisely why GR is a
better theory than the other empirically equivalent ppN ones, otherwise we might have no strong argument against
publishing also, say, post Ptolemaic terms in scientific journals...  It is therefore necessary to define some
other epistemologically relevant measure, besides agreement with the data.  But, which one?

The ability of a theory to {\em predict} nontrivial, yet unobserved, phenomena is rightly considered a strong
evidence of success (see \citealp{HDouglas}, which contains a recent review).  Predictions are certainly invaluable
tools of theory selection, in everyday practice of science.  But, defining precisely what {\em predictions} are,
turns out to be subtler than one might expect.  For instance, it is not too hard to hit a prediction by producing
many possible extensions of an already successful theory.  Are such shots in the dark also 'predictions'?
Predictions are valuable only if their alternatives cannot be equally well justified, which, essentially, leads
again to the necessity of characterizing ad-hoc assumptions, in the first place.

Scientific theories are often evaluated for the opportunities of {\em technological applications} that they promise
to open.  But, either these advantages can be reformulated simply in terms of better empirical adequacy, or, if
not, it is interesting to know {\em why} some theories seem to offer more opportunities than others {\em in spite
  of being empirically equivalent}.  Hence, applications do not answer our question (they are rather one of the
motivations for our question).

One of the most popular tools for theory selection is {\em falsifiability} \citep{Popper}.  But, because of the
Quine-Duhem thesis \citep{Quine2Dogs}, almost no theory can be falsified, as long as any ad-hoc assumption may be
freely added to it.  Therefore, discriminating between the introduction of ad-hoc assumptions and truly new
theories is necessary also to ensure the effectiveness of the criterion of falsifiability.

The idea of {\em reduction} of a theory to a more fundamental one \citep{Nagel}---even if only partially
\citep{KemenyOppenheim} or in some limit \citep{Nowak}---together with the related idea of {\em unification},
singles out essential aspects of true scientific progress.  However, from a logical point of view, nothing prevents
the reducing (or unifying) theory from being an artificial superposition of old theories, made of many and complex
assumptions.  Reductions and unifications represent true progress only if, at the end of the process, some old
assumptions can be dropped.

All this strongly suggests that defining some measure of the amount and/or complexity of the {\em assumptions} does
not only represent a cognitive value in itself, but also a prerequisite for a precise characterization of many
other classic goals of science as well.  The idea is not new.  Many philosophers and scientists (e.g.,
\citealp{Mach,Poincare}, to mention only two of the most influential and modern authors) have stressed the
importance of {\em simplicity, economy of thought} and related concepts\footnote{The previous discussion makes
  clear that what matters, in order to assess the cognitive value of a theory, is always the complexity of its {\em
    assumptions}.  By contrast, the complexity of its {\em consequences} and {\em results} may very well be high,
  which is desirable in a theory that aims at describing the world and its manifest complexity.}.  But, a precise
and general definition is problematic (see e.g., \citealp{Sober_PoS}).  The main obstacle lies in the fact that any
conceivable characterization of simplicity inevitably depends either on the language in which the theory is
formulated, or on some other choice which is equally hard to justify.

A few prominent examples can better clarify this point.  A theory is usually defined as more {\em parsimonious}
\citep{Baker-SEP}\footnote{The review of \citet{Baker-SEP} distinguishes {\em syntactic} from {\em ontological}
  definitions of simplicity.  However, any general definition of simplicity, once it is made precise, it becomes
  syntactic, in some sense.  This is the case also for parsimony.  In this paper, simplicity is always to be
  understood as syntactic simplicity} if it postulates less entities.  But there is no natural and general way to
count the number of entities, and any prescription in this sense inevitably introduces an arbitrary subdivision of
the world into elementary kinds, without convincing justification.  Alternatively, parsimony can be made precise by
identifying the ontological commitment of a theory with the domain of its logical quantifiers \citep{Baker-SEP}.
But this property is not invariant under reformulation of the theory \citep{Quine_OI}.  Another famous definition
of simplicity counts the {\em number of free parameters} that appear in the formulation of the theory
\citep{Popper}.  This is well defined within a fixed class of theories with a fixed common parameterization, but it
becomes arbitrary beyond that.  A further well known example is the proposal of \citet{GoodmanSimple}, that
stimulated much interest and further developments, especially in the 50s and the 60s.  In this case, the complexity
of the theory depends on the choice of the set of {\em primitive predicates}, which is effectively analogous to the
choice of the language \citep{Schwartz_onGoodman}.  Finally, the concept of simplicity derived from {\em Kolmogorov
  complexity} (KC) \citep{Solomonoff,Kolmogorov,Chaitin} has been used by many authors, in recent years, to
determine the so-called universal prior probabilities in a Bayesian context (see
\citealp{LiVitanyi1997,GruenwaldVitanyi2008} for reviews).  It is well known that KC is defined only up to a
constant, that depends on the language.  KC is well suited to study asymptotic properties of theories describing an
increasing amount of empirical data, while keeping the language fixed.  But, KC cannot be used to compare the
simplicity of different theories (each expressed in its own preferred language) with fixed empirical data.  In
fact, for any scientific theory, it is always possible to find a suitable language in which the theory assumes a
trivially simple form \citep{Kelly-razor}.

It should be stressed that the language dependence that characterizes any precise definition of simplicity is not a
problem in itself: an awkward language should obviously produce a complex formulation of the theory.  But, if {\em
  any} theory can be made trivially simple by a suitable choice of the language, then the concept of simplicity
looses any interest.  The idea of simplicity is only meaningful if the {\em simplest} formulation of realistic
theories is {\em not trivial}.  Unfortunately, a common, general argument (hereafter called {\em trivialization}
argument) shows that all previous examples suffer this problem, unless the admissible languages are somehow
limited.  But, how should we justify such limitations?

It is sometimes argued (see, e.g., \citealp{Psillos}, chap. 11) that the special language that can reduce a theory
to a trivial form is artificial and not based on {\em natural kinds}.  This shifts the problem to the one of
characterizing what natural kinds are, which has no convincing solution either \citep{NatKind}.  But there is also
a deeper reason to be skeptical about this approach: one of the main tasks of science is precisely to discover new
kinds (and new languages), which may look weird now, but eventually enable a deeper understanding of the laws of
nature.  The revision of the concept of {\em time} introduced by Einstein and the formulation of particle physics
in terms of {\em quarks} are obvious examples.

In this paper it is stressed that {\em measurability}, rather that {\em naturalness} is the key.  In fact,
scientific theories typically contain concepts that are {\em in principle not measurable}.  Such unmeasurable
concepts should obviously not be used to ground the empirical content of a scientific theory.  Unmeasurable
concepts can certainly be used to formulate the principles of a theory, but then, in order to compute the
complexity of the theory, also the cost of defining the measurable concepts from those used in the principles
should be taken into account.

This idea can be applied to any of the characterizations of simplicity mentioned above.  It should be stressed that
this paper does {\em not} to propose a new notion of complexity, but rather shows how the proper consideration of
the empirical content of a scientific theory prevents a trivialization of essentially any notion of simplicity.
The obstacles preventing trivialization are illustrated in detail with reference to the definition of simplicity
given in Section \ref{sec:concise} ({\em conciseness}).  But the same ideas can be applied to essentially any
acceptable characterization of the simplicity of the assumptions, as discussed in Section \ref{sec:alt}.

The requirement that the formulation of a theory should provide a connection to its measurable concepts may seem
too weak and easy to fulfill.  In fact, as shown in Section \ref{sec:trivial}, this requirement does not rule out
such theories as {\em ``all emeralds are grue''} \citep{grue}, and it also does not offer a solution to the curve
fitting problem (see e.g., \citealp{Sober_PoS}).  But, such {\em toy-models} of scientific theories are only
significant if they capture the relevant features of {\em realistic} theories.  The arguments in Sections
\ref{sec:simple-stab} and \ref{sec:gen-less} show that those models are indeed inadequate.  It is only when the
theory becomes sufficiently rich of consequences that qualitatively new features appear: the connection with
measurable concepts becomes difficult to achieve for those languages that are designed to make most {\em realistic}
scientific theories trivially concise.  In particular, it can be proved that the simple (but not too simple) theory
analyzed in Section \ref{sec:simple-stab} contains unmeasurable concepts.  Moreover, such concepts appear
naturally, when one tries to reformulate the theory in a very concise form.  This provides evidence that the
general trivialization argument reviewed in Section \ref{sec:trivial} is not conclusive, and it also suggests that
the obstacles to trivialization are unlikely to be evaded.

Lacking evidence to the contrary, the fact that some theories can be formulated more concisely than others cannot
be regarded as purely conventional.  Achieving a concise formulation of a realistic scientific theory is far from
easy and highly valuable.

The discussion above makes clear that the notions of simplicity which are significant for science cannot be
properties of the logical or syntactic structure of the theory alone.  Instead, they must depend also on the
connection between the theory and the experience.  For this reason, before examining any concept of simplicity, it
is necessary to define precisely what the {\em empirical content} of a theory is, and what its {\em empirical
  (i.e. measurable) concepts} are.  The traditional approach to these issues is represented by the syntactic {\em
  received} view of scientific theories, originally formulated by the logical empiricists
\citep{CarnapAufbau,Carnap:1958a}.  The main problem with that view is its reliance on a theory-independent
observational language, in order to verify the empirical adequacy of a theory and compare different theories among
each others.  But no such language exists, as it has been convincingly shown by a vast literature (e.g.,
\citealp{Kuhn1,Quine2Dogs,PutnamWTAN,SuppeWW,FraassenImage}).  Perception itself is theory-laden
\citep{Quine_praise} and a self-sufficient phenomenal language is an illusion.  The causal theory of reference for
physical magnitude terms \citep{PutnamER} is often regarded as a way to achieve stability of reference---and hence
enable the comparison of theories with the experience and among each others---in spite of the theory ladenness of
observations.  In this paper, the causal theory of reference is not regarded as a tool to {\em ensure} the
stability of the reference, but rather as a framework to examine {\em under which assumptions} the reference is
sufficiently reliable for the present purposes.  These observations lead to the identification of those syntactic
elements that are necessary to describe the interplay between the empirical content of a theory and its simplicity,
without running into the pitfalls of the received view and while being consistent with the now widely accepted
semantic view \citep{FraassenRep} of theories.  The main message of this paper is that {\em a clear
  identification of the empirical content of a theory lies at the heart of the problem of simplicity}.

The paper is organized as follows.  Section \ref{sec:def-th} introduces those elements of scientific theories which
are needed to provide a relevant characterization of simplicity.  These are further analyzed in Appendix
\ref{app:comm}, in order to show their consistency.  Section \ref{sec:simple} introduces and examines the notion of
conciseness.  In particular, Sections \ref{sec:simple-stab} and \ref{sec:gen-less} show that most realistic
theories cannot be made arbitrarily concise by any known procedure.  Section \ref{sec:alt} extends the previous
result to other definitions of simplicity.  Finally, Section \ref{sec:DDS-TS} examines the possibility that
different definitions of simplicity may converge to produce a consistent characterization of the goals of science.

\section{Scientific Theories And Empirical Concepts}
\label{sec:def-th}

As stressed in the Introduction, in order to provide a characterization of simplicity which is significant for
science, we need to identify precisely a few elements that are part of any scientific theory.  In particular, we
need to specify the role of the {\em principles} and that of the {\em empirical concepts} of a theory.  Similar
concepts occupied the central stage in the traditional syntactic view of scientific theories
\citep{CarnapAufbau,Carnap:1958a,Feigl}, but the latter included unacceptable assumptions that have been the object
of detailed criticisms in the past 50 years, that are briefly reviewed later.  On the other hand, modern semantic
views \citep{SuppesWST,FraassenImage}, concentrate on other aspects of scientific theories (e.g., models), which
are not directly usable for our purposes.  However, \citet{PutnamER} has shown that the empirical concepts ({\em
  physical magnitude terms}) of a scientific theory can be characterized without running into the inconsistencies
of the traditional view.  In this section, we introduce those elements in a way that mimics the received view,
where the latter is unproblematic, but also introduces the crucial corrections dictated by the causal theory of
reference for physical magnitude terms \citep{PutnamER}.  Many comments are postponed to Appendix \ref{app:comm}.
In particular, it is shown in Section \ref{app:sem} that this approach is not inconsistent with a semantic view.

To our purposes, a {\bf scientific theory}, may be viewed as the union of the following elements: a set of abstract
{\em principles}, a set of {\em results}, a set of {\em empirical concepts} and the {\em language} that is used to
express all the previous elements.  The {\bf principles} are abstract, in the sense that they make use of concepts
which are only defined implicitly through the principles themselves.  They merely describe a network of symbols
\citep{Feigl}, and can be seen as a set of mathematical axioms\footnote{In this paper, the words {\em principles,
    postulates, laws, axioms, assumptions and hypotheses} are regarded as equivalent.  No restriction to first
  order logic is assumed.}.  Each theory is regarded as a multidisciplinary collection of principles that include
      {\em all} assumptions (from the logical rules of deduction to the modeling of the experimental devices and of
      the process of human perception) which are needed to derive the results of the theory and compare them with
      the experiments, including a complete estimate of the uncertainties.  All such principles have the same
      epistemological status: even logic rules are to be considered working assumptions and there may be theories
      that adopt different ones.

The {\bf results} comprise all theorems, formulae, rules, solutions of equations, models etc. that have been
derived from the principles of the theory.  The set of results is introduced as a distinct element of the theory,
because its derivation from the principles is not automatic, but requires original intuitions.  Moreover, when a
new theorem is proved, the theory may acquire new empirical consequences and become richer.

The principles and the results are necessarily formulated in some {\bf language}\footnote{Because some languages
  may complicate the comparison with the experiments, as shown in Section \ref{sec:simple-stab}, it is convenient,
  in general, to regard different formulations as different theories.  Nevertheless, we may, for brevity, still
  refer to two different formulations of the same theory, if one is simply the translation of the other in a
  different language.}.  Its terms may be conventionally divided \citep{Feigl}, into {\bf derived concepts}, if
they have an {\em explicit} definition in terms of other concepts of the theory, or {\bf primitive concepts}, if
they are only implicitly defined through the principles.

The {\bf empirical (or measurable) concepts} (ECs) have a double characterization: they are concepts of the theory
(either primitive or derived), and they are also endowed with a set of {\em prototypes} \citep{Rosch}.  The {\bf
  prototypes} of a concept are the subjective examples that a person bears in mind as typical instances of that
concept.  When we need to decide whether a particular phenomenon is an occurrence of a concept or not, we can
compare what we observe to our personal set of prototypes and decide by analogy.  In other words, a prototype for
an EC is a typical member of the {\em extension} of that EC.  Obviously, this does not yet explain how such
prototypes could provide a solid base to science.  This is where the causal theory of reference \citep{PutnamER}
plays a role, but the discussion is postponed to Appendix \ref{app:comm}.

The ECs are further distinguished into {\bf basic empirical concepts} (BECs), that are empirically characterized
(interpreted) {\em only} through a set of prototypes, and {\bf operationally defined empirical concepts} (ODEC),
for whom the theory allows the deduction of a precise operational definition in terms of the BECs\footnote{We are
  not interested in defining the concept of {\em directly} measurable: if---under the assumptions of the
  theory---measuring $A$ implies a definite value of $B$, both $A$ and $B$ are ECs.  It is up to the theory to
  decide which one, if any, is also a BEC.}.  All concepts for which we have neither prototypes nor rules to build
them are {\bf not empirical} (NEC).  The fact that we do not have prototypes or rules associated to a certain
concept does not mean, in general, that it is impossible to build one.  In fact, some NECs may turn out to be ECs
after the discovery of some new experimental technique.  There are, however, also NECs that could not possibly
become ECs.  This crucial observation is discussed in Section \ref{sec:crucial}.

Note, that there is no relation, in general, between primitive concepts and BECs.  The former are related to the
logical structure of the theory, while the latter to the availability of prototypes.  In other words, there is no
obstacle to the existence of primitive-NECs or derived-BECs, as shown in the example in Section \ref{sec:example}.

The division into ECs and NECs evokes the traditional distinction between observational and theoretical terms
\citep{Carnap:1958a}.  However---contrary to Carnap's observational terms---the ECs are theory {\em dependent}.  In
the received view, the observational terms were supposed to represent theory independent sense-data and provided
the basis for radical reductionism and verification theory and also the basis for the comparison of different
theories.  This reconstruction cannot be defended anymore after \citet{Quine2Dogs}\footnote{Note that
  \citet{Quine_praise} himself defends the usefulness of observation sentences, once their theory-ladenness is made
  clear.}: no universal concept can be assumed to be translatable into a purely sense-data language and hence must
be assumed to have a meaning only within some theory.  For this reason, the ECs are here introduced as an
additional label for some theoretical concepts\footnote{Note that the prototypes themselves, like any experiment,
  do not depend on any theory: they are historical events.  But this does not allow to produce theory independent
  BECs, because both the selection and the description of those prototypes that should be relevant to characterize
  a BEC can only be theory dependent.}.  It is of course not obvious how the BECs can enable the comparison of the
empirical statements of different theories.  This is discussed in Appendix \ref{app:compare}.

A different objection \citep{PutnamWTAN,SuppeWW} against the observational-theoretical division deserves special
attention, because it is independent of the theory-ladenness of the observational terms.  \citet{PutnamWTAN} has
observed that there are no terms in the English dictionary that may be regarded as univocally either observational
or theoretical.  For example, the property of being {\em red} can be empirically verified in ordinary objects of
macroscopic size, but its observability is questionable, or certainly impossible, for sufficiently small objects.
\citet{SuppeWW} has further recognized that the observational-theoretical division could be more complex than a
simple bipartition of dictionary {\em terms}, and could involve the context in which the terms are used.  But he
has also argued that such division, if it exists, would be extremely complex, in a way that it is hopeless to
characterize.  These observations are correct: the ECs are not simple dictionary terms.  They include the full
specification of the experimental conditions that the theory considers relevant (and this reinforces their
theoretical dependence).  Moreover, understanding which setup may allow which measurement is the hard and ingenious
work of experimental scientists.  Drawing the complete distinction between the ECs and the NECs would require the
classification of all realizable experimental arrangements where any quantity could be measured.  This is clearly
not feasible.  Moreover, the boundary between ECs and NECs is populated by concepts associated to quantities that
can be measured only with such poor precision that it is questionable whether they are ECs at all.  However, from a
philosophical point of view, a precise and comprehensive compilation of all the ECs is unnecessary: it is
sufficient to recognize that for each scientific theory at least some ECs exist and they can all be constructed on
the basis of both the theory and a small set of BECs.  Only the full list of BECs must be made explicit, as
discussed in Section \ref{sec:simple}.  Also the BECs may not be just dictionary terms: they are rather selected
because of their assumed unambiguity.  For example, most modern scientific theories tend to reduce all the BECs to
the reading of the digital displays of some experimental devices, for which suitable models are assumed.
The classes introduced in this section are summarized in the table below.

\begin{center}
\begin{tabular}{|c|c:c|}
\hline
\multicolumn{3}{|c|}{Concepts of the theory}\\
\hline
\multicolumn{2}{|c:}{ECs}& \multirow{2}{*}{NECs} \\
\cline{1-2}
BECs & ODECs & \\
\hline
\end{tabular}
\end{center}

\subsection{An Example}
\label{sec:example}

Consider, for example, a theory that, besides standard mathematical and logical axioms, also assumes the
Gay-Lussac's law of gases at fixed volume: $P=c T$, where $P$ represents the pressure, $T$ the temperature and $c$
is a constant.  Here, $P$, $T$ and $c$ are primitive concepts.  Let us also assume a suitable model for the
thermometer and the barometer, which can be used, however, only in limited ranges.  As a result $P$ and $T$ are ECs
within those ranges and NECs outside them.  These allow the definition of other ECs such as $c=P/T$, which is hence
a ODEC.  A typical prototype for the EC of $T$ at a reference temperature $T_{\rm ref}\pm\Delta T$ consists in a
sample of real gas equipped with a thermometer that displays the value $T_{\rm ref}$ with a precision of at least
$\Delta T$.  The ECs corresponding to measurements of different temperatures can be characterized by similar
prototypes, but they can also be {\em operationally} defined using the theory (in particular a model for the
thermometer) and a single BEC at the reference temperature $T=T_{\rm ref}\pm\Delta T$.  The choice of the
temperature $T_{\rm ref}$ which is selected as BEC is arbitrary.  But it is important that the {\em necessary}
prototypes can be reduced to those at a single temperature $T=T_{\rm ref}\pm\Delta T$, while all other (measurable)
$T$ correspond to ODECs.

\subsection{A Crucial Property Of The ECs}
\label{sec:crucial}

With no loss of generality, it can be always assumed that the ECs represent properties whose value is either yes or
no.  In fact, any measurement of a real-valued quantity is equivalent to assess whether its value lies or not
within some intervals $[x \pm \Delta x]$, for some $x$ and $\Delta x$.  (Given the limited precision of all
measurements, this is also closer to the experimental praxis.)  In this case {\em a valid prototype should be
  associated to a single connected interval}.  This requirement is necessary to comply with the intuitive idea of
prototype: a single prototype must correspond to a single outcome of a measurement---as inaccurate as it might
be---and not to many precise outcomes at the same time.  If this is not the case for one prototype (e.g. because
the outcome was poorly recorded), a clearer prototype should be provided.  If this is also not possible, one can
only conclude that the corresponding concept is not empirical.

In the example of the previous section, a prototype was represented by an experimental setup where the temperature
of a given sample of gas was measured.  Typically, the thermometer would let us read a number somewhere
between 30.1 \textcelsius\; and 30.2 \textcelsius.  We can accept some uncertainty, which is, in this case
$\sim$0.1 \textcelsius.  Now, imagine that we find a report of the previous day stating that the temperature was
measured once and the result was ``either 29.31$\pm$ 0.01 \textcelsius\; or 32.05$\pm$0.01 \textcelsius''.  We
would conclude that there was a mistake in taking or recording that measurement and we would repeat it.
Experimental results cannot be in macroscopic quantum mechanical superposition states!

This remark plays a central role in this work.  Section \ref{sec:simple-stab} shows that the requirement stated
here---which is indispensable\footnote{Note that this is certainly not a {\em sufficient} condition in order that a
  concept is an EC.}, in order that the ECs have any chance of actually being empirical---cannot be fulfilled by
those very concepts that would naturally make a theory trivially concise.

\subsection{Empirically Equivalent Theories}
\label{sec:EET}

Consistently with the motivations given in the Introduction, we are only interested in considering the relative
simplicity of {\em empirically equivalent} theories.  Empirical equivalence is defined here.

Each scientific theory is motivated by some questions.  A {\bf question} for the theory $T$ consists in the
specification of the values of some concepts of the theory (e.g., the initial conditions or other choices within
the alternatives offered by the principles) and a list of concepts that the theory is expected to determine.  For
example, in astronomy a valid question is: determine the motions of the planets in the sky, knowing the positions
and velocities at some initial time.  It is convenient to distinguish two kinds of questions: {\bf empirical
  questions}, that contain {\em only} ECs, and {\bf technical questions}, that also contain non-empirical concepts
of the theory.  Examples of the latter are questions concerning what cannot be measured in principle, such as the
quantum mechanical wave function, or in practice, because of technical limitations that may be overcome eventually.

Two theories $T$ and $T'$ are said {\bf empirically comparable}, relatively to the sets of ECs ${\cal E}$ of $T$
and ${\cal E}'$ of $T'$, if there is a one-to-one correspondence ${\cal I}$ between ${\cal E}$ and ${\cal E}'$
and---under this correspondence---the experimental outcomes are interpreted in the same way by the two theories,
i.e. those concepts that are identified via ${\cal I}$ possess the same prototypes.  Note that, if $T$ and $T'$ are
comparable for some ECs, then all the empirical questions---limited to those ECs---of one theory are also empirical
questions for the other.  Finally, two theories $T$ and $T'$ are said {\bf empirically equivalent}, relatively to
${\cal E}$ and ${\cal E}'$, if they are comparable and all their results concerning the ECs in ${\cal E}$ and
${\cal E}'$ are equal (within errors) under the correspondence ${\cal I}$.

\section{Simple But Not Trivial}
\label{sec:simple}

This central section shows that there is no reason to expect that realistic theories can be expressed in an
arbitrarily simple form by a suitable choice of the language, while also preserving their empirical content.

First, for the sake of definiteness, a particular definition of simplicity ({\em conciseness}) is introduced in
Section \ref{sec:concise}.  The {\em trivialization} argument, according to which a trivial formulation of any
theory {\em always} exists, is reviewed in Section \ref{sec:trivial}.  But, a gap in the argument is also pointed
out, inasmuch the measurability of the concepts used in the trivial formulation is not granted.  This is not a
remote possibility: in Section \ref{sec:simple-stab} an elementary theory, that involves chaotic phenomena, is
analyzed in detail.  It is actually easy to identify a very concise formulation for it, but precisely those
concepts that naturally enable such trivial formulation can be {\em proved} to be non measurable.  This simple (but
not too simple) theory underlines a serious difficulty in closing the gap of the trivialization argument.

In Section \ref{sec:gen-less} it is stressed that the obstacles identified in Section \ref{sec:simple-stab} are not
due to some very peculiar features of that theory, but they are rather general.  In fact, they are expected to
emerge whenever a theory possesses sufficiently complex consequences.  In view of this, it seems very unlikely that
the gap in the trivialization argument might be closed, for any relevant set of realistic scientific theories.

Finally, Section \ref{sec:alt} considers other possible characterizations of the simplicity of the assumptions,
besides conciseness.  It is shown that any acceptable (as defined below) characterization of the complexity of the
assumptions poses the same obstacles to its trivialization, as conciseness does.

The fact that different characterizations of simplicity are nontrivial does not imply that they are equivalent,
when used for theory selection.  This interesting issue is addressed in Section \ref{sec:DDS-TS}.

\subsection{Definition Of Conciseness}
\label{sec:concise}

Let {\bf $\sigma(T^{(L)})$} denote the {\bf string encoding all the principles} of a theory $T^{(L)}$, where it is
emphasized that the theory is formulated in the language $L$.  As already stressed, it is crucial that the string
$\sigma(T^{(L)})$ include also the definitions of all the BECs in terms of the primitive concepts of the theory.
In this way, anybody able to recognize the BECs of $T^{(L)}$ would find in $\sigma(T^{(L)})$ all the ingredients
that are needed to check\footnote{In order to {\em derive} the results of $T^{(L)}$, the string $\sigma(T^{(L)})$
  is not sufficient, without further original ideas.  However, $\sigma(T^{(L)})$ is sufficient to {\em check} the
  validity of any given derivation. \label{fn:derive}} which results are correctly deduced from $T^{(L)}$, which
questions they answer, and compare them with the experiments.  The {\bf complexity} ${\cal C}(T^{(L)})$ is defined
as the length of $\sigma(T^{(L)})$\footnote{It is interesting to compare this definition with Kolmogorov
  complexity.  The Kolmogorov complexity of a string $x$ is defined as the length of the shortest program written
  in a {\em fixed} Turing-complete language, that outputs $x$.  We could have defined also our complexity as the
  length of the shortest program that outputs the string $\sigma(T^{(L)})$.  However, in the present context, the
  language depends on the theory.  It is therefore equivalent and simpler to define the complexity directly as the
  length of $\sigma(T^{(L)})$, because if we find a shorter program, we can choose that program as
  $\sigma(T^{(L)})$.  Note that $\sigma(T^{(L)})$ is {\em not} expected to produce theorems or formulae
  automatically (see footnote \ref{fn:derive}).  Finally, Kolmogorov theory does not distinguish ECs from NECs,
  although it would not be difficult to introduce an equivalent distinction between realizable and unrealizable
  Turing machines.}\;\footnote{Note that $\sigma(T^{(L)})$ includes all the principles, but not the questions,
  which are potentially unlimited.  However, a theory $T$ cannot cheat by hiding the principles inside the
  questions, because the empirical questions translated from another theory $T'$ through the correspondence ${\cal
    I}$ (see Section \ref{sec:EET}) would miss this information and would have no answer in $T$.}.  The length of
the string is measured in the alphabet associated to the language $L$.  Note that one cannot tell, in general,
whether a given $\sigma(T^{(L)})$ represents the shortest possible formulation of the principles of $T$ in the
language $L$.  The string $\sigma(T^{(L)})$ is simply the shortest {\em known} formulation\footnote{This is
  analogous to the fact that the Kolmogorov complexity function is not computable in general \citep{LiVitanyi1997},
  and most applications of Kolmogorov theory refer to the available compression methods.}  in the language $L$.
The discovery of a shorter encoding represents the discovery of a new result of the theory, enabling a higher
conciseness.  Finally, the {\bf conciseness} of $T^{(L)}$ is defined as the inverse of the complexity ${\cal
  C}(T^{(L)})$.

\subsection{Arguments For The Triviality Of Conciseness}
\label{sec:trivial}

The philosophical literature contains many examples of theories that can be expressed in a very simple form by a
suitable choice of the language.  The classic example is the theory asserting: {\em all emeralds are green if they
  were first observed before January 1st 2020 and blue if first observed after that date} \citep{grue}.  This
statement can be shortened to {\em all emeralds are grue}, by a suitable definition of {\em grue}.  Another example
is provided by the curve fitting problem \citep{Sober_PoS}.  Higher degree polynomials may appear more complex than
lower degree ones, but the complexity disappears under a suitable change of variables.

The concept of conciseness does {\em not} help in deciding which formulation is simpler in these cases.  In fact,
both concepts of green and grue are perfectly measurable and hence acceptable as BECs.  Similarly, high degree
polynomials may look unappealing, but they can be defined and computed precisely in terms of the original
(measurable) variables.  The problem with these toy-models is that they miss some essential features of realistic
scientific theories, insofar as they have very few consequences.  As soon as the theory becomes sufficiently rich
of consequences, qualitatively new obstacles appear, and the path toward a concise {\em and} measurable formulation
is lost, as shown in the example of the next section.

There is also a common {\em general} argument holding that the formulation of {\em any} theory can be made
arbitrarily simple.  In the case of conciseness, such {\bf trivialization argument} goes as follows\footnote{In the
  context of Kolmogorov complexity, the corresponding argument has been presented in
  \citet{Kelly-razor,DelahayeZenil}.}.  Imagine that, in the language $L$, the long string $\sigma(T^{(L)})$ cannot
be compressed further with any known method.  Then one can always define a new language $L'$, which is identical to
$L$ except that it represents the long string $\sigma(T^{(L)})$ with the single character $\Sigma$.  Obviously, it
is impossible to deduce any nontrivial result from a theory whose principles are just '$\Sigma$'.  However, this
might not be necessary, if all the results of $T$ could still be implicit in the {\em interpretation} of $\Sigma$.
In general, one should expect the concept $\Sigma$ to be difficult to interpret in terms of the empirical data.
But the fact that $\Sigma$ may be {\em difficult} to measure is not sufficient to exclude the formulation of the
theory in the language $L'$: difficult measurements can be learned and are routinely conducted by experimental
scientists.

The key point is that there exist concepts that are {\em provably not measurable} (examples are given in the next
section).  In order to be conclusive, the trivialization argument should demonstrate that $\Sigma$ can always be
chosen among the measurable concepts of the theory.  This task has never been undertaken in the
literature\footnote{Remarkably, simplicity and measurability---both classic topics in philosophy of science---have
  been rarely combined.}.  The proof that $\Sigma$ can be chosen---in general---to be measurable is not only
missing, it also looks quite unrealistic.  In fact, the following section illustrates an example of a theory where
the natural choices of $\Sigma$ can be {\em proved} to be unmeasurable.  Alternative choices of $\Sigma$ cannot be
excluded.  But, on the basis of this example, assuming the general existence of measurable $\Sigma$ is definitely
not plausible.

Even if the primitive concepts of the theory are not measurable, it is still possible to define other measurable
concepts and select them as BECs.  In fact, any sentence in the new language $L'$ can still be translated into the
original language $L$ and vice versa.  However, the definition of conciseness requires to take into account also
the length of the string that defines all the BECs in terms of the primitive concepts of the theory.  In the
following example, also this approach is considered, but it happens to lead to lengthier expressions.

\subsection{A Not-too-simple Theory}
\label{sec:simple-stab}

The goal of this section is to show that there exist concepts that are {\em provably not measurable}, and that such
concepts appear naturally when trying to reduce a theory to a trivial form.  This demonstrates a serious gap in the
trivialization argument, which does not ensure that unmeasurable concepts can be avoided.

\begin{figure}[htb]
\centering
\includegraphics[scale=0.30]{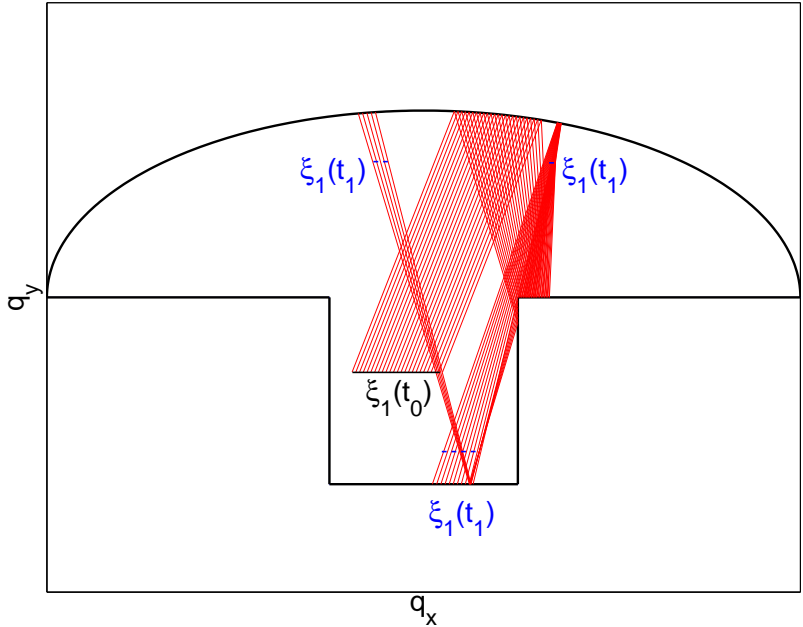}
\caption{The black/solid interval $\xi_1(t_0)$ represents a range of initial conditions in the coordinates $\xi_1$
  at time $t_0$.  This is also an interval in the coordinate $q_x$.  After a few bounces, the interval $\xi_1(t_0)$
  is transformed into at least three disjoint sets (contained in the three blue/dashed lines labeled by
  $\xi_1(t_1)$).  The figure has been produced with the help of the program made available by
  \citet{billmat}. \label{fig:M}}
\end{figure}

To this end, we consider the theory (called ${\cal B}$) which is defined by the laws of classical mechanics applied
to a single small (approximately point-like) ball on a billiard table with a mushroom shape (see, e.g.,
\citealp{MushroomB} and Figure \ref{fig:M}).  This is defined by a curved boundary on the top side (the cap)
joint to a rectangular boundary with sharp corners on the bottom side (the stem).  Such billiards possess chaotic
behaviors, when the initial conditions are chosen within certain values, which are assumed in the following.  The
nice feature of such billiards is that the trajectory of the ball can be computed exactly at any time---in spite of
its chaotic nature.  This enables precise statements about the (non-)measurability of the quantities relevant
to this discussion.

The theory ${\cal B}$ can be naturally expressed in the language $L$ that makes use of the coordinates $z$, where
$z:=(\vec{q},\vec{p})$ denotes together the position $\vec{q}:= (q_x,q_y)$ and momentum $\vec{p}:=(p_x,p_y)$ of the
ball.  The only BEC that needs to be assumed corresponds to assessing, within some fixed precision, whether the
ball at time $t_0$ lies at a reference point $\vec{q}_{\rm ref}$ in the table, and whether it has a reference
momentum $\vec{p}_{\rm ref}$.  Any other measurements of position or momentum (at any time) can be operationally
defined from this single BEC and the principles of the theory.  In fact, the measurement procedures are exactly the
same at any time, since the theory is manifestly time invariant, when expressed in the coordinates $z$ (this does
not hold in the coordinates $\xi$, introduced below).

Measurements of position and time have necessarily limited precision, which is assumed, for definiteness, at the
level of a millimeter and a tenth of a second, respectively.  It is assumed also, for simplicity, that the walls
are perfectly elastic, that the ball does not spin and the friction is negligible for a time sufficient for the
ball to perform a large number of bounces.

Assuming the standard Hamiltonian formalism, the dynamics of this system is completely defined by the function:
$H(z)=H(\vec{q},\vec{p}) = \frac{\vec{p}^2}{2m} + V(\vec{q})$, where $m$ is the mass of the ball, and
$V(\vec{q})=0$ for all $\vec{q}$ inside the billiard and $V(\vec{q})=\infty$ outside.  These formulae contribute to
the length of $\sigma({\cal B}^{(L)})$ with about 35 characters, to which one should add a few more characters to
describe the boundary conditions ($B_M(z)=0$) associated to the mushroom shape of the billiard.  Since the BEC of
this theory ($z_{\rm ref}$) already appears among the primitive concepts, no further definition is needed.
Finally, the contribution to $\sigma({\cal B}^{(L)})$ due to all the standard psychological, physical, mathematical
and logical assumptions, is ignored, since it remains unaltered throughout this discussion.

Following the idea of the trivialization argument, there is a special language ($L'$), that makes the principles of
${\cal B}$ very concise.  The trivialization argument does not explain how to build such a language, nor how to
connect it to measurable quantities.  However, it is not difficult to find a suitable language for the theory
${\cal B}$.  In fact, a natural choice for $L'$ is defined by those coordinates $\xi = (\xi_1, \xi_2, \xi_3,
\xi_4)$, in which Newton's laws take the exceedingly concise form ``$\xi=$ constant''.  Such choice of coordinates
can be defined (with respect to the language $L$) by setting $\xi = \xi(t_0) = z(t_0)$ at a reference time $t_0$
and then assigning the same value of $\xi$ to all future and past configurations that belong to the same trajectory
$z(t)$.

There are now two possibilities.  Imagine, first, that we want to keep the original BEC $z_{\rm ref}=(\vec{q}_{\rm
  ref},\vec{p}_{\rm ref})$.  In this case, the single BEC $z_{\rm ref}$ measured at $t_0$ does not suffice, because
the principles of the theory do not provide the relation between the coordinates $\xi$ and the coordinates $z$ at
any time different from $t_0$.  Hence, we do not know how to perform measurements at times different from $t_0$.
The BECs ($z$) at time $t \neq t_0$ can be related to the primitive concepts $\xi$ by using the Hamiltonian $H(z)$,
the boundary conditions $B_M(z)$, and computing the evolution of the trajectories from $t_0$ to $t$.  These are
computable but very cumbersome expressions, that becomes more and more complex after each bounce.  Since we do not
want to include $H(z)$ and $B_M(z)$ among the principles, such expressions are the only link we have between the
principles and the BECs, and hence we have to include them in $\sigma({\cal B}^{(L')})$, as required by the
definition of Section \ref{sec:concise}.  This implies that $\sigma({\cal B}^{(L')})$ grows indefinitely with the
time separation from $t_0$, while $\sigma({\cal B}^{(L)})$ remains fixed.

The second possibility is to drop the coordinates $z$ altogether, and use the $\xi$ coordinates not only as
primitive concepts in the formulation of the theory, but also directly as BECs.  This leads to a theory that we
denote $\overline{{\cal B}}^{(L')}$, which---apparently---could be much more concise than ${\cal B}^{(L)}$ and yet
empirically equivalent to it.  The problem is that the $\xi$ coordinates, which have a clear interpretation at
reference time $t_0$, cannot be empirically detected at time $t_1$, a few bounces after $t_0$, with the same
precision they were at $t_0$.  This is not just {\em practically difficult} but {\em intrinsically impossible},
because the system ${\cal B}$ displays chaotic dynamics \citep{ChaoticDynamics}, which is characterized by a high
sensitivity to the starting conditions.  This means that two initially nearby trajectories diverge very fast in
time.  To illustrate the consequence of this in a simple way, let us restrict the attention to the two coordinates
$q_x$ and $\xi_1$ of the ball.  By construction, they coincide at $t_0$ (i.e., for any interval at $t_0$,
$[q_x\pm\Delta] = [\xi_1\pm\Delta]$, where $\Delta=1$mm), but at $t_1$ the trajectories that were close at $t_0$
have taken many different directions.  Consequently, the interval $[q_x\pm\Delta]$ at $t_1$ corresponds to many
disjoint and very small intervals\footnote{Because of the sharp (non-differentiable) corners in the boundaries of
  the mushroom shaped billiard, the Poincar\'{e} map---that associates the coordinates of the initial points to
  those of the evolved points---is not continuous.  Hence, a single interval in the parameter set of the initial
  conditions is split, after each bounce, into disjoint intervals.} in the coordinate $\xi_1$.  Conversely, any
interval $[\xi_1\pm\Delta]$ at $t_1$ corresponds to many disjoint and very small intervals in the coordinate $q_x$
(see Figure \ref{fig:M}).  But, there is an important difference between the intervals $[q_x\pm\Delta]$ and
$[\xi_1\pm\Delta]$ at $t_1$: prototypes for the former are possible, while for the latter are not, as a matter of
{\em principle}, because we have no way to measure the many disjoint pieces that compose $[\xi_1\pm\Delta]$.  Of
course, the measurable $[q_x\pm\Delta]$ intervals could be expressed in the $\xi$ coordinates as the union of many
extremely small disjoint intervals, but, as required in Section \ref{sec:def-th}, these cannot be associated to
valid prototypes, and hence the $\xi$ cannot be ECs at $t_1$.  In conclusion, the obvious requirement that ECs are
associated to connected intervals is sufficient to formally exclude---in agreement with the intuition---the $\xi$
concepts as empirical.

In order to use the $\xi$ coordinates to characterize the system at time $t_1$, it would be necessary to introduce
a new coordinates system: besides the $\xi$ with reference at $t_0$, one would need the $\xi^{(t_1)}$, with
reference at $t_1$, and the procedure should be repeated for a full sequence of times $t_i$.  But, the measurements
of $\xi^{(t_i)}$ cannot be operationally defined from those of $\xi^{(t_0)}$, since, as shown in the previous
paragraph, the size of the overlaps of the respective intervals is much below the experimental sensitivity.  Hence,
new BECs---and corresponding new prototypes---are needed for each different time $t_i$.  In order to keep the same
empirical adequacy as the original theory, the new theory should define essentially as many BECs as experimental
data, which would make again $\sigma(\overline{{\cal B}}^{(L')})$ extremely large.

\subsection{Other Scientific Theories}
\label{sec:gen-less}

In the previous section we have examined a particular theory, and showed that the tools at our disposal fail to
make it more concise.  Hence, the theory ${\cal B}$ illustrates some obstacles that prevent closing the gap in the
general trivialization argument of Section \ref{sec:trivial}.  In this section we further note that similar
obstacles appear quite in general for realistic theories.  This should convince the reader that a recovery of some
version of the trivialization argument, covering a relevant set of scientific theories, is very unlikely.

One reason is that, as stressed in Section \ref{sec:def-th}, scientific theories are multidisciplinary collections
of principles gathered from different domains of science.  Because of this, it is sufficient that the mechanism
described in the previous section applies in one corner of the theory, to constrain the possible languages in all
other sectors.  Given that the vast majority of real physical systems admit chaotic phenomena, it is easy to
appreciate the effectiveness of this constraint.  Another reason, which is less compelling but more general, is the
following.  If the laws of a theory are expressed in a form that is so concise that no nontrivial result can be
deduced, then all the consequences of the theory must be evident in the BECs of the theory.  It follows that,
either the theory has very limited consequences, or it needs to introduce a large number of BECs,
or---finally---the interpretation of the BECs is very rich.  But in this last case, it should not be too difficult
to identify not only practical but also {\em fundamental} obstacles to the measurability of those BECs.

It is clear that this argument applies only to theories with sufficiently complex consequences.  Even the idealized
solar system, that played a glorious role in the history of science, is not rich enough---alone---to exhibit the
idea above.  In fact, it may not be impossible to reduce the Ptolemaic model to a very concise theory by using a
small set of suitable BECs.  After all, the orbital motion of a few idealized celestial bodies is an exactly
integrable and periodic system.  But, as soon as one considers, for example, Newton's laws in more general
contexts, the amount and the variety of phenomena that can be described becomes arbitrarily large, while the set of
laws and BECs remains small.  Also the curve fitting problem---which is often employed as a toy-model to discuss
simplicity in the philosophical literature---is not rich enough to show any {\em insuperable} conflict between
conciseness and empirical adequacy, as we have already seen.  Indeed, {\em it is only in a sufficiently rich system
  that the conciseness of the description may come into insurmountable conflict with the accuracy of the
  description}.

This argument is expected to be relevant not only for highly mathematical sciences, but for all theories that
entail many different empirical consequences.  An exhaustive analysis of the implications of this idea for all
scientific fields is obviously impossible here, but one general conclusion can be drawn: for any theory, no trivial
formulation can be assumed to exist (in terms of ECs) unless it is explicitly found.  Hence the available most
concise formulation acquires an objective cognitive value.

\subsection{Nontriviality Of Other Characterizations Of Simplicity}
\label{sec:alt}

In the previous sections we have seen that the trivialization argument fails---in general---to reduce the value of
conciseness, as defined in Section \ref{sec:concise}.  Here, we show that the same result holds for any {\em
  acceptable} definition of the complexity of the assumptions.  In order that a notion of complexity/simplicity be
{\bf acceptable}, we require at least the following two properties.  First, the complexity of a theory should take
into account the cost of defining the BECs of the theory in terms of the concepts appearing in the principles (the
primitive concepts).  Second, the complexity of an expression must be higher than the complexity of any of its
proper sub-expressions\footnote{We also assume that the complexity function takes integer values, so that the
  increments cannot be infinitesimal.}.  These properties are presumably not sufficient to characterize an
acceptable notion of complexity/simplicity, but they are certainly necessary.  These properties hold, in
particular, for our notion of conciseness.  They also hold for the notion of parsimony \citep{Baker-SEP}, which
measures (somehow) the domain of the logical quantifiers that appear in the postulates, or the notion of simplicity
of \citet{GoodmanSimple}, that measures the amount and the complexity of the set of primitive
predicates\footnote{Since the distinction between the BECs and the primitive concepts is usually not stressed, when
  discussing simplicity, the first property is not apparent from \citet{Baker-SEP} and \citet{GoodmanSimple}.  But
  it is obvious, once the definitions of the BECs in terms of the primitive concepts are included among the
  postulates of the theory.}.

If we re-examine the theory of Section \ref{sec:simple-stab}, the same argument goes through unchanged, except for
the points where the complexity of the theories ${\cal B}^{(L')}$ and $\overline{{\cal B}}^{(L')}$ needs to be
computed.  The latter obviously depend on the definition of complexity, but both theories contain expressions that
grow indefinitely with the number of empirical observations to which the theory can be compared.  In fact, the
expressions relating the BECs of ${\cal B}^{(L')}$ to the principle $\Sigma$ become more and more cumbersome, with
increased time separation of the measurement from the reference point.  While, in the case of the theory
$\overline{{\cal B}}^{(L')}$, it is the number of BECs that grows indefinitely with time.  According to the second
requirement stated above, the complexity of a growing expression must grow.  Therefore, we must conclude that none
of those two theories can be simpler than the original theory ${\cal B}^{(L)}$, independently of the particular
definition of complexity which is used.

\section{Different Notions Of Simplicity And The Goals Of Science}
\label{sec:DDS-TS}

In Section \ref{sec:simple} we saw that the general argument for triviality fails, once the empirical content of
the theory is properly taken into account.  Under these conditions, essentially any acceptable characterization of
the simplicity of the assumptions becomes nontrivial.

Thanks to this result, it becomes meaningful to ask whether different characterizations of simplicity also lead to
approximately the same theory selection, when applied to a significant set of real scientific theories.
Furthermore, do they also lead to the same theory selection that may be defined by other classic values in science?
These questions are very important.  The consistency of different criteria would strongly support the high
cognitive value of any such criterion.  Moreover, it would fully justify the scientists' belief that some theories
are unambiguously better than other (empirically equivalent) ones.

Such consistency can never be proved conclusively.  It is only possible to accumulate evidence in its favor or
falsify it\footnote{In this sense, philosophical theories are not different from scientific theories.}.  This can
be done by examining different definitions of simplicity (or different virtues) and applying them to a significant
set of real scientific theories.  Each of these cases clearly requires a dedicated effort, to be duly investigated.
In the rest of this paper we only take a small step in this direction, in order to convince the reader that the
consistency mentioned above is not at all unlikely.

Section \ref{sec:diff} presents a general argument in support of the consistency of criteria based on different
definitions of the simplicity of the assumptions.  As said, this is far from conclusive, but it suggests an
interesting challenge for philosophy of science.  

In the subsequent sections the concept of conciseness is examined in more detail, in order to show that it captures
significant features of the goals of science.  First, in Section \ref{sec:rel-int} it is shown how conciseness can
be estimated in practice.  In Section \ref{sec:ad-hoc} the efficacy of conciseness in penalizing theories with many
ad-hoc assumptions is emphasized.  Section \ref{sec:examples} offers a brief overview of other virtues.

\subsection{Are Different Notions Of Simplicity Equivalent?}
\label{sec:diff}

We have seen that the formulation of a theory must include the definition of its BECs in terms of its primitive
concepts.  Under a different characterization of simplicity, the same theory could achieve its simplest formulation
by using different BECs.  However, the constraints that the BECs should be measurable (ECs) and rather unambiguous
(in order to preserve empirical adequacy) make it {\em very difficult to find formulations that are radically
  different from the traditional one}, which is often already the result of strong efforts of simplification
(according to some intuitive notion of simplicity).  If the choice of the possible formulations is practically
limited to small variations from the traditional one, then the different definitions of complexity must be applied
to the same (or very similar) formulations.  Moreover, we typically want to compare theories that differ only by a
rather limited set of assumptions (see also Section \ref{sec:rel-int}).  These observations together imply that we
typically have to compare different definitions of complexity applied to {\em very similar} and {\em rather short}
strings.  If so, one should expect that simple theories, according to one criterion, be also simple according to
the others, since a short formulation has necessarily also few quantifiers, few predicates, and (except for very
peculiar cases) also the converse is true.  This suggests that all the definitions mentioned in Section
\ref{sec:alt} may lead to essentially the same theory selection, when applied to real cases.

This argument is certainly not conclusive.  It is conceivable that some alternative notion of simplicity might
exist, which is still legitimate and very much different from the intuitive one, and for this reason it might have
been overlooked by the scientists.  It is also possible that the scientists might be overlooking alternative
formulations of their theories that would reveal the prejudices behind their assessments of simplicity.  However,
this can be determined only by providing explicit alternatives and not by general arguments.  In the lack of valid
alternatives, the simplest available formulation retains an objective cognitive value.

\subsection{Practical Estimate Of Conciseness}
\label{sec:rel-int}

The rest of Section \ref{sec:DDS-TS} examines the notion of conciseness and compares it to other classic cognitive
values. The first issue is its practical estimate.

A first remark is that, in order to minimize the conciseness of a theory, it is very hard to use languages that are
radically different from the traditional one.  In fact, this would correspond to a major new discovery.  If we are
limited to small departures from the traditional language, then the conciseness can be estimated by simple
inspection of the length of the principles expressed in their traditional form.

A second remark is that a precise computation of ${\cal C}(T)$ is not realistic, even in a given language, and even
for very simple theories as the one analyzed in Section \ref{sec:simple-stab}.  But, we are never interested in the
absolute value of ${\cal C}(T)$.  The interesting problem, in practice, is always to compare two theories that
share most of the assumptions and are empirically equivalent.  In these cases, the difference ${\cal C}(T)-{\cal
  C}(T')$ between two theories $T$ and $T'$ is typically easy to estimate---possibly using informal languages---and
not impossible to compute exactly.

As an example of how one can estimate the difference ${\cal C}(T)-{\cal C}(T')$ in an informal language, consider
the two theories of special relativity (SR) and classical Galilean relativity (CR)\footnote{Since the two theories
  are not empirically equivalent, the comparison is interesting only from the technical point of view of computing
  their conciseness.}.  In their modern most concise formulations, the two theories differ by a single postulate,
which is, in the case of CR: {\em time and space intervals are constant in all inertial frames}, while for SR it
reads: {\em the speed of light is constant in all inertial frames}.  A suitable language can make these postulates
considerably shorter, but both theories need at least one symbol for each of the concepts of {\em time}, {\em
  space}, {\em interval}, {\em velocity}, {\em light}, etc.  This shows that CR cannot be made more concise than
SR, without a (presently unknown) radical revision of the formulation of these theories. Consequently, if we had to
correct the wrong predictions of CR by adding ad-hoc hypothesis, we would certainly attain a much more complex
formulation than SR.

\subsection{Conciseness, Ad-hoc Assumptions And Information}
\label{sec:ad-hoc}

This section examines the efficacy of conciseness in penalizing theories that include many ad-hoc assumptions.  As
stressed in the Introduction, defining a measure for the amount and complexity of the assumptions is a prerequisite
for a precise characterization of many classic cognitive values in science.  It is well known that the presence of
ad-hoc assumptions is difficult to characterize from a strictly logical point of view.  For example, adding more
assumptions makes a theory more restrictive.  But, the property of being restrictive is not a good characterization
of having many ad-hoc assumptions, because the best theories are extremely restrictive and admit only what really
happens.  What is bad in ad-hoc assumptions is not that they introduce restrictions, but that we are unable to
express them without adding {\em more words}, while a good theory manages to be very restrictive with few words.
Consideration of the syntax, besides the logical structure, is clearly necessary to represent the intuitive idea of
ad-hoc assumptions.  If the shortest formulation of the theory $T'$ is not longer than the one of $T$, then $T'$
cannot be seen as the addition of ad-hoc assumptions on top of $T$, even if $T'$ implies $T$.  This means that a
{\em nontrivial} measure of conciseness can---at least in some cases---exclude that a new theory is obtained by
adding ad-hoc assumptions.

For example, most theories of gravity within the ppN formalism \citep{lrr-2006-3} are build as modification of
Einstein's (or Newton's) theory of gravity.  For many of those ppN theories, we cannot imagine a way to express
them more concisely than Einstein's (Newton's) theory itself: we know how to formulate them only by formulating
Einstein's (Newton's) theory first, and then adding further elements.  Moreover, under the reasonable assumption
that the Lagrangian formalism and differential geometry are standard tools which are needed anyway, it is hard to
imagine a theory as concise as general relativity and empirically equivalent to it.  These ppN theories are
generally recognized as possessing more ad-hoc assumptions than general relativity, and they actually correspond to
longer formulations.

Another example is the following.  Assuming that a given thermometer was not working properly on some specific
occasions, may explain a few strange results.  But, if we try to explain all strange measurements of temperature in
this way, we have to add to the general theory a huge list of specific exceptions.  Alternatively, assuming that
all thermometers of some brand give a wrong answer 10\% of the times can contribute to provide a more consistent
description of a large set of measurements around the world, with a limited increase of the complexity of the
theory.  The latter procedure is clearly less ad-hoc and more concise than the former.

The sensitivity to ad-hoc assumptions is a consequence of a more basic virtue of concise theories: out of two
theories with the same consequences, the more concise one provides evidence that {\em the needed information to
  obtain the same results is less} than it would be expected from less concise theories.  This is also confirmed by
the following observation.  If a scientist is confronted with two formulations of the same theory she would never
erase the shorter one, even if it misses other qualities that she might find desirable.  In that case she would
keep both.  This highlights an important {\em cognitive advantage} of the most concise formulation, which is
completely independent from any reference to reality, and hence fits well an {\em empiricist} view of science.

\subsection{Conciseness And The Goals Of Science}
\label{sec:examples}

This section sketches some connections between the concept of conciseness and other classical criteria for
theory appraisal.  Again, this is not meant to show any superiority of conciseness with respect to other
characterizations of simplicity, but rather to exemplify how a well defined and nontrivial characterization of
simplicity gives the chance to establish explicit connections with other cognitive values.

As already mentioned in the introduction, the idea of conciseness may enable a more precise formulation of the idea
of {\em unification}.  Two theories are unified when they are substituted by a single theory that answers at least
all the questions previously answered by the original two theories.  If the unification is not mere juxtaposition,
some of the old assumptions should appear as duplicated and be combined in a single one, or be both dropped in
favor of another more powerful assumption.\footnote{It may be the case that a unifying theory introduces more
  sophisticated mathematical tools.  But, according to the definition in Section \ref{sec:def-th}, a scientific
  theory is necessarily a multidisciplinary collection of assumptions coming from different fields.  Sophisticated
  mathematical tools---besides being generally very concise---have usually many fields of applicability, which
  considerably reduce their impact in the overall conciseness.}  This suggests that most interesting cases of
unification have also produced more concise theories, although a systematic historical analysis would be certainly
needed to assess this point conclusively.

A similar argument can be used to interpret many cases of {\em reduction} of scientific theories as cases of
increased conciseness\footnote{Note that, if $T$ is more empirically adequate than $T'$, it is not very interesting
  to compare the conciseness of $T$ to the one of $T'$, but rather to the one of a theory $T''$, which is obtained
  by adding to $T'$ suitable assumptions able to correct the wrong predictions of $T'$. }.  Classic examples are
Newton's reduction of Kepler's laws to the laws of mechanics, and the reduction of thermodynamics to statistical
mechanics\footnote{It is controversial whether the latter is an example of reduction, but it is anyway an example
  of increased conciseness.}.  In the first case, the laws that describe mechanical phenomena are shown to be
sufficient also to explain astronomical phenomena; in the second case the laws of mechanics and probability are
sufficient to explain also thermodynamical phenomena.  Both cases correspond to the realization that all the
phenomena under consideration can be explained with less overall assumptions.  Other examples are being provided,
currently, by computational sciences, that have achieved tremendous successes in reducing various phenomenological
laws to more fundamental ones.

Among the recognized values, that a scientific theory should have, is also that of {\em coherence} with the other
accepted theories.  This does not seem to be related to conciseness.  But, in our approach (see Section
\ref{sec:def-th}), a scientific theory is necessarily a multidisciplinary collection of {\em all} the assumptions
that are needed to derive the results that can be compared to real experiments.  In this context, coherence between
the different domains of science is not a virtue: it is a necessity, that is assumed at the start.

An original application is the explanation of the problem of {\em fine tuning} in the standard model of elementary
particles.  This problem lies in the fact that the fundamental parameters of the model need to be known with a very
large number of digits, in order to reproduce (even with moderate precision) the experimental values.  Since the
fundamental parameters must be included in the principles of the theory, this is, effectively, a problem of
conciseness.

The idea of conciseness can also explain why {\em solipsism} is void of interest.  Solipsism cannot be excluded
neither on logical nor on empirical grounds.  The problem with solipsism is rather the unnecessary amount of
assumptions that need to be made in order to explain the experience.  In fact, the experiences {\em reported} to
the subject by other people require different explanations---and hence additional postulates---from those
explaining the {\em direct} experiences of the subject.  What the subject sees can be explained much more {\em
  concisely} by assuming a underlying reality, independent of the mind.

Finally, one should also mention that there exist research programs that aim at recognizing signatures of {\em
  irreducible complexity} in nature.  In such programs, conciseness cannot be a value, by construction.  But, this
is consistent with the fact that those goals are not recognized by the vast majority of the scientific community,
since no evidence can possibly exclude the existence of yet uncovered more concise rules.

\section{Conclusions And Perspectives}
\label{sec:conclusions}

Scientists often regard simpler assumptions as unambiguously preferable to complex ones.  Moreover, most classic
standards of progress in science implicitly rely on a characterization of the simplicity of the assumptions, in
order to acquire a precise meaning.

Any precise definition of simplicity---which is relevant in this sense---necessarily requires the examination of
the principles of the theory.  Moreover, in order to evade general arguments for the triviality of any notion of
simplicity, it is also necessary to establish a formal connection between the principles and the measurable
concepts of the theory (ECs).  This paper shows explicitly how the principles and the ECs can be included in a view
of scientific theories, which is not in contradiction with modern views, and avoids the pitfalls of the traditional
view.  Although the ECs are, in general, theory dependent, each theory includes concepts that are empirical by
construction of the theory itself.

The ECs are important not only in order to compare the theory with the experiments and with other theories, but
also to constraint its possible formulations.  In fact, {\em a theory must be expressed in a language able to
  represent its empirical content}.  The importance of this requirement cannot be appreciated when considering
isolated toy-theories, that entail only few consequences.  But it becomes crucial for realistic theories, whose
consequences are many and complex.  In fact, in those cases, improving the simplicity of the formulation may
conflict with the need of preserving its accuracy.  This is illustrated through the inspection of a specific
example of a theory and by employing the precise notion of conciseness.  As a result, {\em the fact that some
  theories are more concise than others is not purely conventional.  It is as objective as the fact that some
  quantities are measurable and others are not}.

The concept of conciseness introduced in this paper is just one of the many possible characterization of
simplicity.  Here it is used mainly as an example, showing that a nontrivial characterization of simplicity is
possible.  Similar arguments can be applied to other definitions of simplicity, that also become nontrivial, once
the precise connection to measurable quantities is taken into account.

These observations lead naturally to the important question whether different---nontrivial---definitions of
simplicity induce approximatively the same theory selection, when applied to a significant set of real cases.  A
further question is whether these criteria are also consistent with the other classic standard of progress in
science.  The availability of a class of nontrivial definitions of simplicity is a crucial pre-requisite to address
these questions precisely.  {\em A positive answer to these questions would provide a solid philosophical
  justification on support of the scientists' belief that some theories are unambiguously better than other
  (empirically equivalent) ones}.  This paper cannot support a positive answer conclusively, but it argues, through
a few general considerations and some examples, that this possibility is not presently excluded.  In order to prove
the scientists wrong, it is necessary to identify a legitimate definition of simplicity that contradicts some of
the assessments that are universally held by the scientists.  General arguments about its existence are not
sufficient.

\paragraph{\bf Acknowledgments}
Email exchanges with L.D.~Beklemishev, G.~Chaitin, M.~Hutter and H.~Zenil, as well as the valuables comments of the
anonymous referees are gratefully acknowledged.  The author is a member of the Interdisciplinary Laboratory for
Computational Science (LISC) and acknowledges support from the AuroraScience project.

\appendix
\section{Appendix. Representing And Comparing The Empirical Contents}
\label{app:comm}

The main goal of this paper is to illustrate how the empirical content of a theory represents an obstacle to the
simplicity that the theory can attain.  In Section \ref{sec:def-th} it was proposed to describe the empirical
content of a theory through the ECs.  This appendix shows that the ECs are actually well suited for this purpose,
in particular they enable the comparison of the empirical statements of different scientific theories.  This is not
obvious, since the ECs are theory dependent, and the prototypes are subjective.  The key ideas are those of
\citet{PutnamER}.

Semantic incommensurability \citep{Feyerabend62,Kuhn1} notoriously challenges the possibility of comparing the
statements of two different theories\footnote{Methodological incommensurability \citep{Bird-SEP}, which refers to
  the incomparability of the {\em cognitive values} used to judge different theories, is not considered in this
  appendix.  But the conclusions of this paper clearly support the general value of simplicity, next to that of
  empirical adequacy.}.  Semantic incommensurability may be further distinguished \citep{Scheffler2} into {\em
  variation of sense} and {\em referential discontinuity}.  Variation of sense refers to the difficulty of
interdefining concepts that are implicitly defined within different axiomatic systems.  This problem has been
investigated within, e.g., the structuralist program \citep{structuralism-SEP,structuralism}.  Interdefinability is
only possible after establishing logical relations of reduction between theories, which are possible only in very
limited cases.  This is not sufficient for our goals.

Referential discontinuity, on the other hand, corresponds to the problem that the ECs of different theories may
fail to refer to the same phenomena.  Referential continuity is sufficient to ensure the existence of a common
ground for the comparison of two scientific theories and the lack of interdefinability is not an obstacle to it.

There are other obstacles to referential continuity.  One problem is that different people may have different
interpretations of the BECs, since these are based on sets of subjective prototypes.  This is considered in Section
\ref{app:basic}.  A second problem is that two scientific theories may use different ECs and it is not clear how
the comparison of the empirical adequacy is possible.  This is the subject of Section \ref{app:compare}.  Finally,
Section \ref{app:sem} shows that this view is consistent with the semantic one.

\subsection{The Relation Between Prototypes And BECs}
\label{app:basic}

The relation between prototypes and BECs can be established by pointing our finger to prototypes and by naming
them.  Everyone taking part to such ceremonies of baptism will then extend, by analogy, the set of her own
subjective prototypes.  This mechanism is the basis of the causal theory of reference for physical magnitude terms
\citep{PutnamER}, which is a special case of the causal theory of reference for natural kinds
\citep{Kripke,Putnam}.  This theory has been extensively studied and its limits are well known.  In particular,
when we point our finger and assert: ``that is the symbol {\tt 0} on the display'', misunderstandings can {\em
  never} be excluded completely.  Even if some people agree that some prototypes are in the extension of a BEC, we
can never be sure that the same people will always agree on the assignment of new prototypes that may eventually
appear in future.  Each ambiguity can be eliminated by agreeing on further prototypes, but other ambiguities are
always possible \citep{Quine_indet}\footnote{The causal-descriptive theory of reference (see e.g.,
  \citealp{Psillos}, chap. 12) has also been proposed as a way to constrain the possible interpretations.  But, on
  one hand, the BECs have a formal definition in terms of the primitive concepts of the theory, which may be
  regarded as a descriptive element.  On the other hand, no description can ensure a stable reference, even in
  combination with ostension \citep{Quine_indet}.}.  In fact, the procedures of baptism establish {\em
  correlations} between the subjective extensions that different people associate to the same BECs---which are
partially tested by the feedback that these people return to each other---but they can never {\em guarantee a
  one-to-one correspondence} between these sets.  However, correlations are precisely all what science needs.
Referential discontinuities, like those envisaged here, can be seen as some of the many unavoidable sources of
experimental errors.  What is necessary for a scientific theory is not to eliminate them, but to produce an
estimate of the probability and magnitude with which they occur.  For example, the agreement of different observers
on those BECs, whose prototypes are simple pictures, holds with high probability (that can be estimated) under the
assumption of a neuropsychological theory connecting the light signals hitting the retina with the formation of
pictures in the mind, that can be classified by analogy, and the assumption that such mechanisms are rather similar
across humans.  The inclusion of these assumptions increases the complexity of the theory (see Section
\ref{sec:simple}), but also protects it from being ruled out by a single experimental oversight.

These assumptions have the same epistemological status as the other principles of the theory and all face together
the tribunal of experience \citep{QuineNat}.  (There cannot be truly {\em non-problematic} assumptions in a modern
view of scientific theories.)  This makes it harder to identify the assumptions which are responsible for a bad
matching between a theoretical prediction and the empirical data.  But this is a practical and not an
epistemological problem.

In order to maximize the probability of correlation between the different subjective interpretations of the BECs, a
scientific theory has the possibility to select a convenient small set of BECs.  Modern theories, in particular,
tend to reduce every BEC to the reading of the digital displays of suitable experimental devices, for which
appropriate theoretical models are assumed\footnote{There may still be people who stubbornly refuse to see the
  difference between the digits {\tt 0} and {\tt 1} printed on a display.  There is no way to {\em prove} them
  wrong---without assuming other BECs---if they insist that this is what they see, but there are sociological
  theories that tell how often such eccentric behaviors may appear.}.  As a result, modern scientific theories
predict a strong (and quantifiable) correlation between the subjective extensions that different people assign to
the same BEC.  These theories effectively associate each of their BEC to an approximately well identified and
observer-independent set of prototypes.

If a theory cannot quantify the correlation between the subjective extensions of its own BECs, this is a problem
for the theory, that presumably has a poor empirical adequacy, but not an epistemological problem.  In conclusion,
{\em when a scientific theory properly includes all the theoretical assumptions which are necessary to predict the
  experimental results, a failure of reference is not a problem for the theory of reference (nor for epistemology):
  it is a potential problem for the scientific theory itself}.  This conclusion is completely consistent with the
idea of naturalized epistemology \citep{QuineNat} for which this section represents nothing more than a concrete
exemplification.

\subsection{The Relation Between The ECs Of Different Theories}
\label{app:compare}

After having identified the assumptions supporting a stable interpretation of the ECs of a single theory, it is
necessary to consider the relation between the ECs of different theories.  In order to compare their empirical
statements, two scientific theories $T$ and $T'$ must provide two sets of ECs ${\cal E}$ and ${\cal E}'$ that can
be identified through a correspondence ${\cal I}$, as required in Section \ref{sec:EET}.  The ECs ${\cal E}$ and
${\cal E}'$ do not have to share the same {\em meaning}---which may not be possible between concepts belonging to
incompatible theories \citep{Feyerabend62}---but should have the same {\em extension} (\citealp{Hempel}, p. 103),
i.e., in particular, share the same prototypes.  In other words, the two theories should stipulate coinciding
measuring {\em procedures}, even though the {\em justifications} and the {\em descriptions} of such procedures
could be very different.

It is natural to ask whether it is always possible to find two non empty sets ${\cal E}$ and ${\cal E}'$ with these
properties.  Answering this question in full generality is well beyond the goals of this paper.  For the present
purposes, it is enough to remark that this is possible for a wide class of real scientific theories whose
comparison is interesting.  This is actually the case even for classic examples of syntactic incommensurability.
For instance, the concept of {\em mass} in Newtonian mechanics and the one in relativistic mechanics have different
meanings and extensions.  But {\em there exist procedures to measure space-time coordinates}, which are valid for
both theories (once the reference frame is specified), and the numerical results are in one-to-one correspondence.
This means that the corresponding ECs have the same extension.  Moreover, there seem to be no examples of real
scientific theories whose comparison would be interesting but it is impossible because of lack of corresponding ECs
(\citealp{GodfreySmith}, p. 92), and even those authors that defend the relevance of syntactic incommensurability
for real science (e.g., \citealp{Carrier}) insist that this is only meaningful for theories that do identify some
of their ECs.

Note that the set ${\cal E}$, supplied by $T$ to establish a comparison with $T'$, does not enable a comparison
also with any other theory.  If, later, we want to compare $T$ with another theory $T''$, we may need to choose a
new set of ECs $\hat{{\cal E}}$, among those that can be defined within $T$.  For example, measurements of absolute
time represented legitimate ECs for classical dynamics, and were associated to prototypes where the speed of the
clock was disregarded.  Such ECs enabled the comparison of pre-relativistic theories, but are not suitable for
comparing those theories with special relativity.  In fact, as a consequence of the theory-ladenness of the ECs,
there is no lingua franca of observations: the empirical languages change with the emergence of new theories, but
this does not typically hinder their comparability.

Sometimes, the kind of problems described in Section \ref{app:basic} are revealed by the introduction of a new
theory.  For example, the theory of special relativity reveals that measurements of absolute time may give
inconsistent results, if taken from different reference frames.  However, if an EC (e.g., absolute time), which is
expected to be sufficiently unambiguous according to the theory $T$ (Galilean relativity), turns out to be
ambiguous after the introduction of $T'$ (special relativity), this is a problem for the empirical adequacy of the
assumptions of $T$, rather than a problem for the comparison of $T$ with $T'$.  In other words, once we have
identified two sets of ECs ${\cal E}$ and ${\cal E}'$, {\em the conditions for the comparability of $T$ with $T'$
  are already part of the assumptions that both $T$ and $T'$ need to incorporate, in order to formulate their own
  experimental predictions.}

\subsection{Syntactic And Semantic Views}
\label{app:sem}

The discussion in Section \ref{sec:def-th} is formulated in a language that bears many similarities with the one
used within the received syntactic view of scientific theories.  Although we have stressed the crucial differences,
it is also worth comparing with modern semantic approaches \citep{SuppesWST,FraassenImage}.  Here it is shown that
the present view differs from the one of \citet{FraassenRep} only in the {\em emphasis} on some syntactic
aspects---which are necessary for the purposes of this paper---and it is otherwise consistent with it.

In any semantic approach a central role is played by the possible {\em models} of a scientific theory
\citep{SuppesWST,FraassenImage}.  Models are, in general, a combination of results derived from the principles of
the theory (e.g., a class of solutions of Newton's differential equations of motion) together with specific initial
conditions.  Models contain original informations with respect to the principles, because (as stressed in Section
\ref{sec:def-th}) the derivation of the results typically requires original ideas.  Moreover, models can be used
directly to produce theoretical predictions and to perform a comparison with the experiments.  It is true that many
properties of a scientific theory can be conveniently appreciated by examining its set of models, and consideration
of its axiomatic structure is unnecessary for that.  But, models typically include many {\em consequences} of the
theory, besides its minimal set of {\em assumptions}.  For this reason, they are not suitable to evaluate the
complexity of the assumptions, which is our goal.  This is more directly expressed by the principles of the theory.
Note that, according to Section \ref{sec:def-th}, the principles really include everything that is necessary to
derive measurable predictions.  For example, if a general theory admits different solutions, and if the measurable
initial conditions (which are part of the questions) do not allow the complete identification of the relevant
solution, suitable assumptions must be added to the principles, in order to select a single solution.

It is interesting to pursue the parallel with \citet{FraassenRep} somewhat further.  In particular, van Fraassen's
{\em empirical substructures} can be identified with those results of the theory that can be expressed exclusively
in terms of ECs.  Furthermore, \citet{FraassenRep} emphasizes the fact that measurements are {\em representations},
that need the specification of the context and the experimental setup, in order to allow the interpretation of the
outcome.  As stressed in Section \ref{sec:def-th}, all these informations must be part of the ECs.  Note that the
compatibility of the two views is possible because both van Fraassen's empirical substructures and the ECs
introduced here are integral parts of the theory, and not above it.

The connection with the phenomena is achieved, in both the present and van Fraassen's approach, via indexicality
(and the identification of prototypes).  As stressed by \citet{FraassenRep}, indexical statements plays a central
role also in evading Putnam's paradox \citep{PutnamRTH}, which states that almost any theory can be seen as a true
theory of the world.  In fact, it is almost always possible to find a correspondence between the concepts of a
theory and the phenomena in the world, such that---with this {\em interpretation}---the theory is true.  In order
to evade this paradox, one needs to fix the correspondence between the ECs and the phenomena in an independent way.
Such independent constraints are imposed precisely by indexical statements (with all the caveats already
explained): when we point our finger, we insist that {\em this} is the symbol {\tt 0} on the display and not
whatever suits the theory in order that the theory is correct.

\bibliography{../philo}{}
\bibliographystyle{chicago}
\end{document}